\documentclass[aps, prl, twocolumn,reprint,superscriptaddress,longbibliography]{revtex4-1}

\usepackage[utf8]{inputenc}
\usepackage[T2A,T1]{fontenc}
\usepackage[full]{textcomp}
\usepackage[dvipsnames]{xcolor}
\usepackage{bm}
\usepackage{graphicx}
\usepackage{float}
\usepackage{wrapfig}
\usepackage{placeins}

\usepackage{booktabs}
\usepackage{multirow}
\usepackage{subcaption}

\usepackage{amsmath}
\usepackage{gensymb}
\usepackage{nicefrac}
\usepackage{chemformula}
\usepackage{natbib}
\DeclareSymbolFont{cyrillic}{T2A}{cmr}{m}{n}
\DeclareMathSymbol{\Sha}{\mathalpha}{cyrillic}{216}
\usepackage{braket}
\usepackage{dcolumn}
\usepackage{txfonts}
\usepackage[classicReIm]{kpfonts}
\usepackage{mathdots}
\usepackage{enumitem}
\usepackage{soul}
\usepackage{tabularx}
\usepackage{booktabs}
\usepackage{multirow}
\usepackage{siunitx}

\usepackage[linktocpage=true,colorlinks=true,pdfborder={0 0 0},linkcolor=blue,citecolor=blue,filecolor=yellow,urlcolor=blue,bookmarks,pdfauthor={},]{hyperref}

\newcommand{\Brown}{Department of Physics \&{} Astronomy, Brown University, USA}
\newcommand{\BCTPI}{Brown Center for Theoretical Physics and Innovation (BCTPI), Brown University, USA}
\newcommand{\Imperial}{Blackett Laboratory, Imperial College London, UK}
\newcommand{\CERN}{Experimental Physics Department, European Organization for Nuclear Research (CERN), Switzerland}
\begin{document}

\title{Contrastive Metric Learning for Point Cloud Segmentation in Highly Granular Detectors}


\author{Max Marriott-Clarke}
\affiliation{\Imperial}
\affiliation{\CERN}
\author{Lazar Novakovic}
\affiliation{\Brown}
\author{Elizabeth Ratzer}
\affiliation{\Imperial}
\author{Robert J. Bainbridge}
\affiliation{\Imperial}
\author{Loukas Gouskos}
\affiliation{\Brown} \affiliation{\BCTPI}
\author{Benedikt Maier}
\affiliation{\Imperial}

\begin{abstract}
We propose a novel clustering approach for point-cloud segmentation based on supervised contrastive metric learning (CML). Rather than predicting cluster assignments or object-centric variables, the method learns a latent representation in which points belonging to the same object are embedded nearby while unrelated points are separated. Clusters are then reconstructed using a density-based readout in the learned metric space, decoupling representation learning from cluster formation and enabling flexible inference. The approach is evaluated on simulated data from a highly granular calorimeter, where the task is to separate highly overlapping particle showers represented as sets of calorimeter hits. A direct comparison with object condensation (OC) is performed using identical graph neural network backbones and equal latent dimensionality, isolating the effect of the learning objective. The CML method produces a more stable and separable embedding geometry for both electromagnetic and hadronic particle showers, leading to improved local neighbourhood consistency, a more reliable separation of overlapping showers, and better generalization when extrapolating to unseen multiplicities and energies. This translates directly into higher reconstruction efficiency and purity, particularly in high-multiplicity regimes, as well as improved energy resolution. In mixed-particle environments, CML maintains strong performance, suggesting stable learning of the shower topology across particle types, while OC exhibits significant degradation. These results demonstrate that similarity-based representation learning combined with density-based aggregation is a promising alternative to object-centric approaches for point cloud segmentation in highly granular detectors.
\end{abstract}

\maketitle

\section{Introduction}

Modern particle detectors increasingly rely on high-granularity sensor technologies that provide detailed spatial and timing measurements of energy deposits. The resulting data naturally form point clouds with irregular geometry and variable size. A central reconstruction task is the segmentation of these point clouds into groups of measurements originating from individual particles. In high-granularity calorimeters such as the CMS High Granularity Calorimeter (HGCAL)~\cite{CERN-LHCC-2017-023}, this task is particularly challenging due to the frequent spatial and energetic overlap of particle showers, requiring algorithms capable of resolving complex and highly correlated hit patterns.

Graph neural networks (GNNs) have emerged as a powerful framework for point-cloud reconstruction in particle physics~\cite{GNN_Particle,Topol_gnn,GNN_track}. 
A widely used learned clustering approach built on this framework is object condensation (OC)~\cite{oc}, in which the network predicts object-centric latent variables that guide cluster formation.
By integrating clustering into the training objective, OC achieves strong performance across a range of reconstruction tasks~\cite{oc_GNN,oc_qasim,oc_track,oc}. However, this formulation tightly couples representation learning to a specific clustering procedure. In dense environments, where multiple nearby showers compete for representative points, this coupling can lead to ambiguities in both the learned clustering coordinates and the assignment of hits to objects.

In this work, we propose an alternative clustering paradigm based on contrastive metric learning (CML)~\cite{Dim_reduction}. Rather than predicting object-level variables, the method learns a latent embedding in which hits from the same particle shower are placed nearby while hits from different showers are separated. Clustering is performed only after training and acts as a readout of the learned representation. This decoupling allows the embedding geometry to be optimized directly for pairwise compatibility, without imposing constraints associated with a particular clustering mechanism.

A key advantage of this formulation is that the contrastive objective depends only on relative relationships between hits, rather than absolute object-level properties~\cite{Dim_reduction}. In object-centric approaches, the network must learn representative points and clustering scales that are implicitly tied to the detailed morphology of particle showers. As a result, mis-modelling in simulation can propagate directly into the learned clustering variables. In contrast, CML requires only that hits from the same shower be more similar than those from different showers, which may make the learned representation less sensitive to variations in shower shape, energy response, and event composition, though we do not demonstrate this empirically in the present work.

To extract clusters from the learned embedding, we introduce a density-based readout procedure tailored to metric spaces. This method identifies representative points from local neighbourhood structure and forms clusters without requiring explicit object-level predictions, ensuring consistency with the geometry induced by the contrastive objective.

The proposed approach is evaluated on simulated electromagnetic (EM) and hadronic (HAD) showers in a detector model inspired by the CMS HGCAL. A direct comparison with OC is performed using identical GNN backbones and matched embedding dimensionality to isolate the impact of the learning objective. Representation quality is assessed using embedding geometry metrics, while reconstruction performance is evaluated using physics-based observables including efficiency, purity, and energy resolution.

We find that CML produces a more structured embedding geometry, with improved separability particularly in the tails of the distance distributions where reconstruction failures typically occur. This leads to significant improvements in clustering performance in high-multiplicity environments with strong shower overlap. These results demonstrate that learning a similarity-based representation provides a robust alternative to object-centric clustering methods for high-granularity detectors and more generally for dense point-cloud segmentation problems.

\section{Methods}

\subsection{Model Architecture}
\label{subsec:model_architecture}

Both learning objectives employ an identical GNN backbone to permit a direct comparison. Differences between methods are restricted to the task-specific output heads required by the respective loss functions. 

Each calorimeter hit is represented by a five-dimensional input feature vector
\[
(x, y, z, E, L),
\]
where $(x,y,z)$ denote the hit position, $E$ is the deposited energy, and $L$ is the detector layer index. 
Further details of the dataset and input representation are given in Section~\ref{sec:Experimetnal_setup}. 
Events are treated as variable-size point clouds with no graph connectivity provided as input. Truth shower identifiers are stored separately and are used only for defining the learning objectives and evaluation metrics. No preprocessing or feature normalisation is applied to the inputs, and we find this has negligible impact on performance.

The features are first projected into a learned latent space using a two-layer multilayer perceptron (MLP) with hidden dimension 64 and exponential linear unit (ELU) activations, producing a 64-dimensional representation.

The encoded hit features are subsequently processed by three DynamicEdgeConv layers~\cite{dgcnn}, each constructed using a $k$-nearest-neighbour graph ($k=24$) in the current latent feature space using Euclidean distance. A learnable edge function is applied and aggregated using the permutation-invariant "max" operation. Because the graph connectivity is recomputed after every DynamicEdgeConv layer, the neighbourhood structure evolves during training, allowing the network to jointly learn feature representations and a data-driven similarity metric between hits. Constructing the graph in latent feature space reduces sensitivity to the geometric constraints imposed by the detector layout and enables interactions between hits across layers, which is important for capturing the longitudinal development of calorimeter showers.

After the shared DynamicEdgeConv backbone, the network separates into task-specific output heads. In both cases, features are processed by a two-layer MLP (64$\rightarrow$32) with ELU activations and dropout ($p=0.1$), applied identically across both heads. For the contrastive objective, a final linear layer projects the 32-dimensional features to a 16-dimensional embedding used for metric learning. For object condensation, two parallel linear heads are applied to the same 32-dimensional features, predicting a per-hit condensation score $\beta_{i} \in (0,1)$ and 16-dimensional clustering coordinates $c_i$. The embedding dimensionality is matched between the two methods to ensure a comparable latent space dimensionality. Both methods were also evaluated with a reduced four-dimensional embedding; the corresponding results are presented in Section~\ref{sec:results}.

Since all backbone layers (the input MLP, the three DynamicEdgeConv layers, and the shared output MLP), activation functions, and optimization hyperparameters are shared, differences in embedding structure and reconstruction performance between the two methods can be primarily attributed to the choice of learning objective. The two methods are trained entirely independently; the shared backbone refers to identical architectural design only, not joint optimization.

\subsection{Contrastive Metric Learning}

CML is a class of methods~\cite{cpc, simclr, supcon} that learns a representation space in which samples sharing the same semantic identity are embedded nearby, while unrelated samples are separated. Rather than predicting discrete labels or cluster assignments, the objective directly shapes the geometry of the latent space, making the embedding itself the primary target of optimization. This paradigm is well suited to point-cloud segmentation in high-granularity detectors, where reconstruction depends on identifying compatible sets of hits rather than assigning them to predefined object templates. Such objectives have been widely used in representation learning and clustering tasks~\cite{simclr, cont_rep_rev, cont_clust}, where downstream performance depends on the separability of learned features.

We apply supervised contrastive learning at the hit level within individual events. For an event containing $N$ hits with embeddings $\{\mathbf{z}_i\}_{i=1}^{N}$ and corresponding shower labels $\{y_i\}_{i=1}^{N}$, positive pairs are defined as hits originating from the same simulated particle shower, while negative pairs correspond to hits from different showers within the same event. Comparisons across events are excluded since separate events do not share physically meaningful relationships.

The embeddings are $\ell_2$-normalized and compared using cosine similarity $s_{ij} = \mathbf{z}_i^{\top}\mathbf{z}_j$, constraining all representations to lie on the unit hypersphere. This encourages hits from the same shower to form compact regions in angular space while separating hits from different showers, thereby learning a discriminative similarity metric without requiring explicit cluster centres or assignment variables. This formulation is advantageous in calorimeter reconstruction, where showers are spatially extended and frequently overlap, making object boundaries inherently ambiguous. Since the DynamicEdgeConv graph is constructed in the current latent feature space at each layer, the neighbourhood structure used for message passing evolves alongside the learned representation. For CML, this creates a coupling between the contrastive objective and the graph topology: as the embedding learns to place shower-compatible hits nearby, the graph increasingly connects hits from the same shower, reinforcing the representation. This feedback between representation learning and graph construction may contribute to the stability of the CML embeddings observed in Section~\ref{sec:results}.

The loss is given by the supervised contrastive (SupCon) objective~\cite{supcon}:
\begin{equation}
\mathcal{L}_{\mathrm{SupCon}} =
- \frac{1}{|\mathcal{A}|}
\sum_{i \in \mathcal{A}}
\frac{1}{|\mathcal{P}(i)|}
\sum_{p \in \mathcal{P}(i)}
\log
\frac{
\exp\!\left( \mathbf{z}_i^\top \mathbf{z}_p / \tau \right)
}{
\sum\limits_{j \neq i}
\exp\!\left( \mathbf{z}_i^\top \mathbf{z}_j / \tau \right)
},
\end{equation}
where each hit $i$ acts as an anchor (i.e., a reference point against which all other hits in the event are compared), $\mathcal{A}$ denotes the set of anchors with at least one positive sample, $\mathcal{P}(i)$ is the set of positive hits for anchor $i$ (i.e., hits sharing the same shower label), and $\tau$ is a temperature parameter that controls the concentration of the embedding distribution on the hypersphere. Smaller $\tau$ sharpens the similarity distribution and encourages tighter clusters. In this work we use $\tau=0.1$. The 16-dimensional output of the final linear projection layer constitutes the embedding used for all metric learning and evaluation; no intermediate representation is used.

Rather than sampling pairs, all hits within the event contribute to the loss: positive pairs appear explicitly in the numerator, while all negative pairs contribute through the partition function in the denominator, actively shaping the gradient. Calorimeter events typically contain many weak negatives and relatively few informative hard negatives, such that random sampling often yields low-information pairs. Aggregating over all pairs ensures that informative relationships consistently contribute to the optimization. As shown in the SupCon formulation~\cite{supcon}, the gradient naturally emphasises hard positives and hard negatives, providing implicit hard pair mining. Although contrastive methods often rely on large batch sizes to obtain sufficient negatives, the large number of hits within each event naturally provides a rich negative set, allowing each event to act as an effective training batch.

\subsection{Object Condensation Baseline}

The OC~\cite{oc} is a learned clustering method that reconstructs particle candidates from detector hits using per-hit latent variables. The approach has been applied across a range of high-energy physics tasks~\cite{oc, oc_qasim, oc_track, oc_tracking}, and is adopted here as a baseline.

For each hit $i$, the network predicts clustering coordinates $\mathbf{c}_i \in \mathbb{R}^{d}$ and a condensation score $\beta_i \in (0,1)$. The coordinates define a latent clustering space, while $\beta_i$ identifies representative hits. Hits with large $\beta$ act as condensation points around which compatible hits are grouped. In this work, the clustering space dimension is $d=16$, matching the contrastive embedding dimension to ensure a comparable latent space dimensionality. 

The condensation score $\beta$ is mapped to a positive charge 
\begin{equation}
q_i = \operatorname{arctanh}^2(\beta_i) + q_{\min},
\end{equation}
which determines the interaction strength in the clustering space. For each object $k$, the hit with the largest charge defines the condensation point
\begin{equation}
\alpha_k = \arg\max_{i \in k} q_i,
\end{equation}
where $k$ indexes the set of hits belonging to a given truth shower and $K$ denotes the total number of showers in the event.

Hits belonging to the same object are attracted to the condensation point, 
\begin{equation}
\mathcal{L}_{\mathrm{att}} =
\frac{1}{K}
\sum_{k=1}^{K}
\frac{1}{|k|}
\sum_{i \in k}
q_i q_{\alpha_k}
\left\lVert \mathbf{c}_i - \mathbf{c}_{\alpha_k} \right\rVert^2 ,
\end{equation}
while hits from different objects are repelled within a finite radius,
\begin{equation}
\mathcal{L}_{\mathrm{rep}} =
\frac{1}{K}
\sum_{k=1}^{K}
\frac{1}{|\bar{k}|}
\sum_{i \notin k}
q_i q_{\alpha_k}
\max\!\left(0,\, 1 - \left\lVert \mathbf{c}_i - \mathbf{c}_{\alpha_k} \right\rVert \right).
\end{equation}

A regularisation term encourages condensation points to attain high confidence,

\begin{equation}
\mathcal{L}_{\beta} =
\frac{1}{K}
\sum_{k=1}^{K}
\left(1 - \beta_{\alpha_k}\right).
\end{equation}

The total event loss is
\begin{equation}
\mathcal{L}_{\mathrm{OC}} =
s_{\mathrm{att}}\mathcal{L}_{\mathrm{att}} +
s_{\mathrm{rep}}\mathcal{L}_{\mathrm{rep}} +
s_{\mathrm{coward}}\mathcal{L}_{\beta}.
\end{equation}

In our implementation $s_{\mathrm{att}} = s_{\mathrm{rep}} = s_{\mathrm{coward}} = 1.0$ 
and $q_{\min} = 0.1$, following the default configuration of~\cite{oc}. A limited scan of the loss 
weights was performed and found to produce negligible changes in reconstruction performance relative to the default configuration. The model learns latent variables that define clustering behaviour, while the actual grouping of hits is performed at inference time using these learned quantities.

For direct comparison, the OC model uses the same DynamicEdgeConv backbone described in Section~\ref{subsec:model_architecture}, differing only in the output heads predicting $\beta_i$ and $\mathbf{c}_i$. Training is performed event-wise with all hits assigned to their corresponding truth shower, consistent with the supervised setting used for contrastive learning. In contrast to CML, which learns a global similarity metric, OC directly optimizes object-centric clustering variables in the latent space. In this formulation, clustering structure is explicitly encoded through object-level attractors rather than emerging from pairwise representations.

\subsection{Clustering}
Clustering is performed after training and treated as a readout of the learned representation rather than part of the learning objective. All clustering is performed independently for each event.

\paragraph{Agglomerative clustering.}
As a common baseline, agglomerative clustering~\cite{Agg} is applied directly to the learned embeddings produced by both CML and OC. We use Ward linkage with a Euclidean distance metric applied to the $\ell_2$-normalised embeddings; on the unit hypersphere, Euclidean and cosine distances are monotonically related. The method does not require a fixed number of clusters and naturally accommodates events with variable particle multiplicity. The number of clusters is determined by applying a distance threshold $\delta_{\mathrm{agg}}$ to the Ward linkage dendrogram: clusters separated by a distance greater than $\delta_{\mathrm{agg}}$ are not merged. The value of $\delta_{\mathrm{agg}}$ is optimized on the auxiliary dataset and reported in Table~\ref{tab:clustering_thresholds}. Applying the same clustering algorithm to both embeddings enables comparison of the intrinsic quality of the learned representations, independent of method-specific inference procedures.

\paragraph{Object condensation inference.}
For OC, clusters are obtained using the inference procedure associated with the OC loss~\cite{oc}. Candidate condensation points are selected by thresholding the predicted scores $\beta_{\text{OC},i} > t_\beta$ and sorting in decreasing $\beta_{\text{OC}}$. A greedy separation step enforces a minimum Euclidean distance $t_d$ between selected points in the learned clustering coordinate space $\mathbf{c}_i \in \mathbb{R}^{16}$. Hits within radius $t_d$ of a condensation point are assigned to it, and any remaining hits are assigned to the nearest condensation point. The values of $t_\beta$ and $t_d$ are given in Table~\ref{tab:clustering_thresholds}.

\paragraph{Density-based readout of metric embeddings.}
Contrastive embeddings do not predict representative points. We therefore define a clustering readout operating directly in the embedding space, deriving candidate centres from local neighbourhood structure rather than from a learned score.

For each hit, a local density estimate is obtained from the distance to its $k$-th nearest neighbour $d_k(i)$. In this work we use the same value of $k$ for the density estimate as for the DynamicEdgeConv graph construction during training, although these two choices are conceptually independent. This distance is mapped to a score
\begin{equation}
\beta_{\text{CML},i} = \exp\!\left(-\frac{d_k(i)}{\tau}\right),
\end{equation}
where $\tau$\ is the temperature parameter used in the contrastive loss. This mapping is consistent with the similarity scaling used during training, assigning large scores to densely populated regions of the embedding.

Candidate centres are selected by thresholding $\beta_{\text{CML},i} > t_\beta$ and enforcing a minimum separation $t_d$. Clusters are formed by assigning hits within distance $t_d$ to each centre, followed by nearest-centre assignment for remaining hits. Here $\beta_{\text{CML},i}$ plays a role analogous to the condensation score in OC, but is computed from the embedding rather than predicted by a network head.

Unlike standard density-based methods such as DBSCAN~\cite{dbscan} and HDBSCAN~\cite{McInnes2017}, which rely on fixed density thresholds, the proposed readout operates on local neighbourhood structure in the learned embedding, making it compatible with spatially varying density across showers.

\paragraph{Summary.}
Three clustering strategies are evaluated: shared agglomerative clustering applied to both embedding spaces, the native OC inference procedure, and the proposed density-based readout for contrastive embeddings. This design isolates the effect of the learned representation from the choice of clustering algorithm.

\section{Experimental Setup}
\label{sec:Experimetnal_setup}

\subsection{Dataset}

The datasets used in this study are produced with a standalone Geant4~\cite{GEANT} simulation closely resembling the HGCAL detector~\cite{HGCAL_TD}.

The calorimeter model comprises 50 longitudinal layers: 28 electromagnetic (CE-E) followed by 22 hadronic (CE-H). The transverse cell size is $1\,\mathrm{cm}\times1\,\mathrm{cm}$ over a $100\,\mathrm{cm}\times100\,\mathrm{cm}$ area, ensuring full shower containment. Energy deposits are calibrated in minimum-ionising-particle (MIP) units, with the simulated hit response consistent with test-beam observations~\cite{HGCALTB_E,HGCALTB_Pion}.

Three training datasets are considered: an electromagnetic (EM, electrons) sample, a hadronic (HAD, charged pions) sample, and a mixed sample containing both particle types. The EM dataset contains 2–10 electrons per event with primary energies uniformly distributed between 30 and 400\,GeV. The HAD dataset contains 2–7 pions per event in the same energy range. The mixed dataset contains an equal fraction of electrons and charged pions with 2–7 particles per event and primary energies between 30 and 250\,GeV. For each configuration, $200\,000$ events are used for training and $50\,000$ for validation.

Models are evaluated on independent test datasets containing between 1 and 30 particles per event with primary energies ranging from 30 to 600\,GeV. Separate EM and HAD test samples are used. Models trained on the EM and HAD datasets are evaluated on their respective particle types, while the model trained on the mixed dataset is evaluated on both.

In addition to the training, validation, and primary test datasets, an auxiliary optimization dataset is generated for each training configuration with the same multiplicity and energy distributions as the corresponding training sample. These events are not used during network training and are used only to determine the clustering hyperparameters at inference time. 
Optimal clustering thresholds are selected by maximising reconstruction performance on this dataset. For models trained on the EM and HAD samples, thresholds are determined separately. For the mixed model, a single set of thresholds is determined on the mixed optimization dataset and applied to both test samples. The resulting threshold values are summarised in Table~\ref{tab:clustering_thresholds}.

\begin{table}[htbp]
\centering
\caption{Clustering thresholds. The parameter $\delta_{\mathrm{agg}}$ denotes the agglomerative clustering distance threshold, while $t_{\beta}$ and $t_d$ correspond to thresholds used in both the density-based readout and object condensation inference.}
\label{tab:clustering_thresholds}
\begin{tabular}{llcccc}
\toprule
Model & Method & $\delta_{\mathrm{agg}}$ & $t_{\beta}$ & $t_d$ \\
\midrule
\multirow{2}{*}{EM}
    & CML & 8 & 0.4 & 0.45 \\
    & OC  & 6.75 & 0.1 & 0.25 \\
\midrule
\multirow{2}{*}{HAD}
    & CML & 10 & 0.4 & 0.45 \\
    & OC  & 11 & 0.1 & 0.35 \\
\midrule
\multirow{2}{*}{Mixed}
    & CML & 9.5 & 0.4 & 0.45 \\
    & OC  & 7.5 & 0.1 & 0.3 \\
\bottomrule
\end{tabular}
\end{table}

When multiple particles deposit energy in the same cell, their contributions are merged into a single hit, as in realistic detector readout. Each hit is assigned the label of the particle contributing the largest fraction of its deposited energy. The fractional contribution of the dominant particle is recorded to define hit purity, which is close to unity across the dataset. This quantity is not used as an input feature.

All particles are generated within a cone of fixed opening angle, with a half-angle of $5^\circ$ around the detector axis. Consequently, increasing the event multiplicity leads to a systematic increase in local particle density within the calorimeter volume. This provides a controlled mechanism to vary the degree of shower overlap, allowing reconstruction performance to be probed across regimes of increasing ambiguity and hit-level confusion.

Table~\ref{tab:xyz_metrics} reports the average number of hits per event and the degree of spatial overlap between showers in raw detector coordinates, computed using the same neighbourhood metrics defined in Section~\ref{sec:embed_metrics} but applied directly to the $(x,y,z)$ hit positions rather than learned embeddings. The point clouds are large: at $N=2$, EM events contain on average $2\,180$ hits and HAD events $3\,148$ hits, rising to $26\,493$ and $43\,661$ respectively at $N=30$, reflecting the broader lateral and longitudinal extent of hadronic showers. The spatial overlap metrics confirm that the reconstruction task is intrinsically challenging. At $N=30$, the nearest-neighbour recall R@1 drops to $0.711$ for EM and $0.509$ for HAD in detector space, with contamination C@10 reaching $0.334$ and $0.549$ respectively, indicating that for hadronic showers at high multiplicity roughly half of the ten nearest spatial neighbours of a given hit originate from a different shower. These values motivate the need for a learned representation that can disentangle showers that are not separable in input space.

\begin{table}[htbp]
\centering
\caption{Spatial overlap metrics and average hit counts computed in 
detector coordinate space $(x,y,z)$ at representative multiplicities. 
Metrics are defined in Section~\ref{sec:embed_metrics} and applied 
directly to the hit positions without any learned transformation. 
The large point-cloud sizes and high contamination values at 
$N=30$ illustrate the intrinsic difficulty of the segmentation task.}
\label{tab:xyz_metrics}
\begin{tabular}{c c c c c r}
\toprule
Dataset & $N_{\rm particles}$ & $R@1$ & $R@10$ & $C@10$ 
  & $\langle N_{\rm hits} \rangle$ \\
\midrule
\multirow{3}{*}{EM}
& 2  & 0.980 & 0.996 & 0.022 & 2\,180 \\
& 15 & 0.784 & 0.953 & 0.246 & 15\,406 \\
& 30 & 0.711 & 0.924 & 0.334 & 26\,493 \\
\midrule
\multirow{3}{*}{HAD}
& 2  & 0.942 & 0.992 & 0.072 & 3\,148 \\
& 15 & 0.629 & 0.910 & 0.426 & 24\,461 \\
& 30 & 0.509 & 0.846 & 0.549 & 43\,661 \\
\bottomrule
\end{tabular}
\end{table}

\subsection{Training Details}

All models are trained using the Adam optimizer~\cite{adam} with an initial learning rate of $3\times10^{-4}$ and a batch size of 16 events. The learning rate is reduced by a factor of 0.5 every 25 epochs, and training is performed for up to 100 epochs. Identical optimization settings, data splits, and initialization procedures are used for both CML and OC models, with differences arising only from the learning objective and corresponding output heads.

Model selection is based on the validation loss evaluated after each epoch on an independent validation set. The checkpoint with the lowest validation loss is retained, and early stopping is applied once no further improvement is observed.

Training is performed using mixed-precision arithmetic with gradient scaling. All models are trained on a single GPU (NVIDIA Tesla V100, 32\,GB, or NVIDIA H100, 80\,GB), with consistent results observed across both hardware configurations.

\section{Evaluation}

\subsection{Embedding Geometry Metrics}
\label{sec:embed_metrics}

The learned representations are evaluated independently of any clustering procedure by quantifying geometric properties of the embedding space within each event. Such metrics are commonly used to assess representation quality in metric learning settings~\cite{DML_Spher, ML_reality}. All quantities are computed using cosine similarity on $\ell_2$-normalized embeddings and averaged over events.

\paragraph{Recall@$k$ (R@$k$)}
For each hit, the $k$ nearest neighbours in embedding space are identified. A hit is considered correctly retrieved if at least one neighbour originates from the same truth shower. The fraction of such hits defines R@$k$. We report R@1 and R@10, which probe whether compatible hits are placed as the closest neighbour and within the local neighbourhood, respectively.

\paragraph{Contamination@10 (C@10).}
Local neighbourhood purity is quantified as the fraction of the 10 nearest neighbours that originate from a different truth shower. Unlike Recall@$k$, which provides a binary criterion, this metric captures the continuous fraction of incorrect neighbours. Lower contamination indicates that local neighbourhoods correspond more closely to physically consistent showers.

\paragraph{Pairwise separability.}
Positive (same-shower) and negative (different-shower) hit pairs are sampled and their cosine similarities compared. The area under the receiver operating characteristic (AUC) curve measures the probability that a positive pair has higher similarity than a negative pair, providing a global measure of separability in the embedding space.

\paragraph{Intra- and inter-shower distances.}
To characterise the geometric structure of individual showers, the analysis is restricted to the energetic core of each shower by retaining the highest-energy hits that together account for $95\%$ of the total shower energy. This suppresses low-energy peripheral deposits while preserving the physically relevant structure.

Cosine distances $d = 1 - \mathbf{z}_i^\top\mathbf{z}_j$ are computed between hits. Intra-shower compactness is quantified by the $99^{\mathrm{th}}$ percentile of same-shower pairwise distances, denoted $d^{Q99}_{\mathrm{intra}}$, capturing the largest separations within a shower core. Inter-shower separation is measured as the median distance between hits from different showers, providing a robust estimate of typical separation between distinct objects.

\subsection{Reconstruction Metrics}
\label{sec:ReconMetrics}

The performance of reconstructed particle candidates is evaluated using standard high-energy physics (HEP) metrics following the CMS TICL validation procedure~\cite{Ticlv5}. These quantities assess the physical quality of reconstructed objects rather than the learned representation.

\paragraph{Efficiency.}
A simulated particle is considered reconstructed if at least $70\%$ of its deposited energy is contained within a single reconstructed object. The reconstruction efficiency is defined as
\[
\varepsilon = \frac{N_{\mathrm{matched\ sim}}}{N_{\mathrm{sim}}}.
\]

\paragraph{Purity.}
Purity quantifies the extent to which a reconstructed object originates from a single simulated particle. This is evaluated using the Reco-to-Sim score $S(r \rightarrow s)$, defined for a reconstructed object $r$ and simulated particle $s$ as
\begin{equation}
S(r \rightarrow s) =
\frac{\sum\limits_{\mathrm{hits}}
\max\!\left(0,\, f_{\mathrm{reco}} - f_{\mathrm{sim}}\right)^{2} E^{2}}
{\sum\limits_{\mathrm{hits}} f_{\mathrm{reco}}^{2} E^{2}},
\end{equation}
where $f_{\mathrm{reco}}$ and $f_{\mathrm{sim}}$ are the fractional energy contributions of a hit to the reconstructed object and simulated particle, respectively, and $E$ is the hit energy. The score is zero for a perfectly pure object and increases with contamination. A reconstructed object is considered pure if there exists at least one simulated particle $s$ for which $S(r \rightarrow s) < 0.2$. The purity is then defined as
\[
P = \frac{N_{\mathrm{pure\ reco}}}{N_{\mathrm{reco}}}.
\]

\paragraph{Multiplicity ratio.}
To quantify the tendency of the clustering to split or merge objects, the ratio
\[
R_N = \frac{N_{\mathrm{reco}}}{N_{\mathrm{sim}}}
\]
is used, where $R_N > 1$ indicates object splitting and $R_N < 1$ indicates merging.

\paragraph{Energy resolution.}
For each simulated particle, the reconstructed object with the largest shared energy is selected as the match. Objects with $S(r \rightarrow s) > 0.5$ are discarded as impure. The energy response is defined as
\[
r = \frac{E_{\mathrm{reco}}}{E_{\mathrm{sim}}},
\]
where $E_{\mathrm{reco}}$ is the sum of the energies of the hits assigned to the reconstructed object and $E_{\mathrm{sim}}$ is the true particle energy.

Within bins of $E_{\mathrm{sim}}$, the mean response $\mu$ is used to define a calibrated response $r' = r / \mu$. The energy resolution is then defined as the effective sigma $\sigma_{\mathrm{eff}}$, given by half the minimum interval containing $68.3\%$ of the $r'$ distribution. For Gaussian distributions, $\sigma_{\mathrm{eff}}$ is equivalent to the standard deviation, while remaining robust to non-Gaussian tails and outliers~\cite{Sigma_eff}.

As a reference, we also compute the resolution of an ideal pattern-recognition algorithm that perfectly assigns all hits to their truth shower, with no merging or splitting. The energy of each reconstructed object is taken as the sum of the energies of all hits assigned to the corresponding truth shower. This provides a lower bound on the achievable resolution given the detector response and the hit-level energy calibration, and is independent of any clustering algorithm.

\section{Results}
\label{sec:results}

\subsection{Embedding Geometry}

\begin{figure*}[t]
    \centering
    \includegraphics[width=1\linewidth]{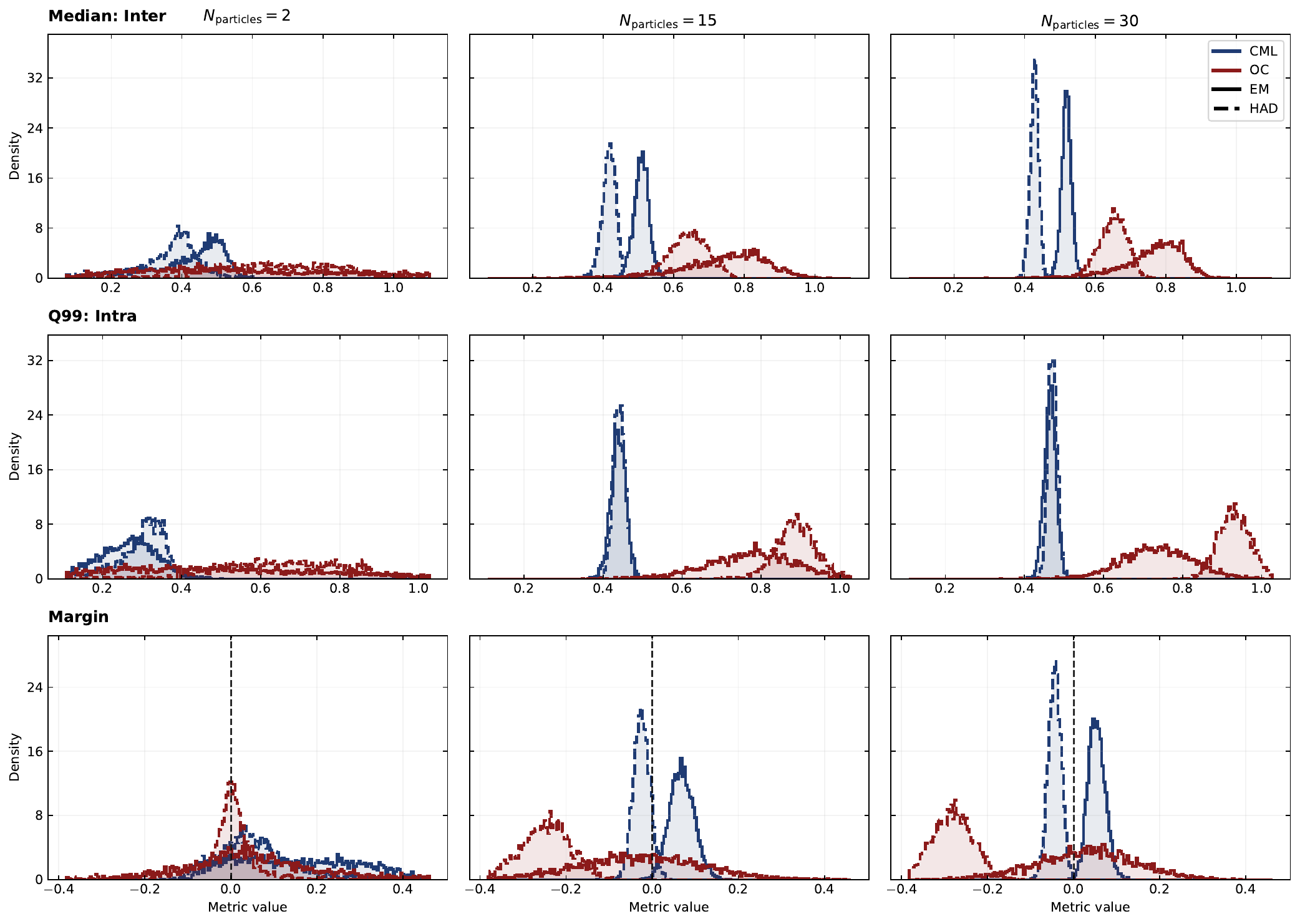}
    \caption{Event-level embedding-geometry distributions for EM and HAD showers at increasing generated particle multiplicity ($N_{\mathrm{particles}}$). Top: median inter-shower separation $d_{\mathrm{inter}}^{\mathrm{median}}$. Middle: intra-shower distance tail $d_{\mathrm{intra}}^{\mathrm{Q99}}$. Bottom: separation margin $\Delta = d_{\mathrm{inter}}^{\mathrm{median}} - d_{\mathrm{intra}}^{\mathrm{Q99}}$. Narrow positive or near-zero margins indicate a well-defined clustering scale, while broad or negative margins indicate increasing ambiguity between showers.}
    \label{fig:embedding_quality}
\end{figure*}

Embedding-level performance is evaluated separately for the EM and HAD datasets using the metrics defined in Section~\ref{sec:embed_metrics}. Results are reported at representative event multiplicities to probe behaviour as shower overlap increases.

Across both datasets, CML consistently outperforms OC, achieving higher Recall@$k$, lower contamination@10, and higher AUC at all tested multiplicities (Table~\ref{tab:embed_metrics}). The performance gap is modest at low multiplicity but increases steadily with shower density, indicating that CML preserves more stable local neighbourhood structure under increasing overlap.

\begin{table}[t]
\centering
\footnotesize
\setlength{\tabcolsep}{3pt}
\renewcommand{\arraystretch}{1.05}

\caption{Embedding-geometry metrics for EM and HAD showers at representative multiplicities. CML consistently achieves higher Recall@$k$, lower contamination@10, and higher AUC than OC, with the performance gap increasing as multiplicity increases.}
\label{tab:embed_metrics}

\begin{tabular}{c c c cc c c}
\toprule
Dataset & $N_{\rm particles}$ & Method
& \multicolumn{2}{c}{$R@k$}
& $C@10$
& AUC \\
\cmidrule(lr){4-5}
& & & $1$ & $10$ & & \\
\midrule

\multirow{6}{*}{EM}

& \multirow{2}{*}{2}
& CML & \textbf{0.941} & \textbf{0.992} & \textbf{0.067} & \textbf{0.956} \\
&  & OC  & 0.935 & 0.989 & 0.072 & 0.944 \\

& \multirow{2}{*}{15}
& CML & \textbf{0.691} & \textbf{0.932} & \textbf{0.339} & \textbf{0.950} \\
&  & OC  & 0.667 & 0.920 & 0.363 & 0.923 \\

& \multirow{2}{*}{30}
& CML & \textbf{0.572} & \textbf{0.886} & \textbf{0.467} & \textbf{0.945} \\
&  & OC  & 0.536 & 0.871 & 0.503 & 0.918 \\

\midrule

\multirow{6}{*}{HAD}

& \multirow{2}{*}{2}
& CML & \textbf{0.938} & \textbf{0.992} & \textbf{0.077} & \textbf{0.945} \\
&  & OC  & 0.922 & 0.989 & 0.092 & 0.927 \\

& \multirow{2}{*}{10}
& CML & \textbf{0.744} & \textbf{0.943} & \textbf{0.286} & \textbf{0.927} \\
&  & OC  & 0.698 & 0.927 & 0.330 & 0.889 \\

& \multirow{2}{*}{20}
& CML & \textbf{0.628} & \textbf{0.900} & \textbf{0.407} & \textbf{0.917} \\
&  & OC  & 0.573 & 0.876 & 0.460 & 0.877 \\

\bottomrule
\end{tabular}
\end{table}

Figure~\ref{fig:embedding_quality} shows the corresponding embedding-geometry distributions. The top row presents the median inter-shower separation $d_{\mathrm{inter}}^{\mathrm{median}}$, while the middle row shows the intra-shower distance tail $d_{\mathrm{intra}}^{\mathrm{Q99}}$. For CML, both quantities remain narrowly distributed across all multiplicities, indicating a stable embedding geometry with consistent distance scales both within and between showers. In contrast, OC produces much broader distributions, reflecting substantially larger event-to-event variability.

The bottom row shows the separation margin 

\[
\Delta = d_{\mathrm{inter}}^{\mathrm{median}} - d_{\mathrm{intra}}^{\mathrm{Q99}},
\]
which directly measures whether showers remain geometrically separable. At $N=2$, the margin is broad for both methods, as with only two showers per event absolute distances are unconstrained; however, the CML margin remains centred at positive values, confirming separability even in this regime. At higher multiplicities, where genuine shower overlap is present, CML maintains a narrow margin distribution with a positive peak for EM showers and a near-zero peak for HAD showers, indicating a well-defined and stable clustering scale for both particle types.

In contrast, OC produces broad and often negative margin distributions. For EM showers, the margin spans both positive and negative values, implying strong overlap between intra- and inter-shower distances. For HAD showers, the margin is systematically negative and becomes increasingly broad with multiplicity. In both cases, no single distance threshold can separate showers consistently across events.

Overall, the advantage of CML arises primarily from controlling the tails of the distance distributions rather than shifting their central values. By maintaining compact intra-shower structure together with stable inter-shower separation, CML produces embeddings that remain reliably clusterable even in dense environments.

\subsection{Reconstruction Performance}

Reconstruction performance is evaluated using the physics metrics defined in Section~\ref{sec:ReconMetrics}. Across all datasets and reconstruction configurations, CML outperforms OC, with the largest gains appearing at high multiplicity. These differences follow directly from the embedding geometry: the narrow margin distributions of CML define a stable clustering scale, whereas the broad and often negative margins of OC make clustering increasingly ambiguous.

\begin{figure*}[t]
    \centering
    \includegraphics[width=1\linewidth]{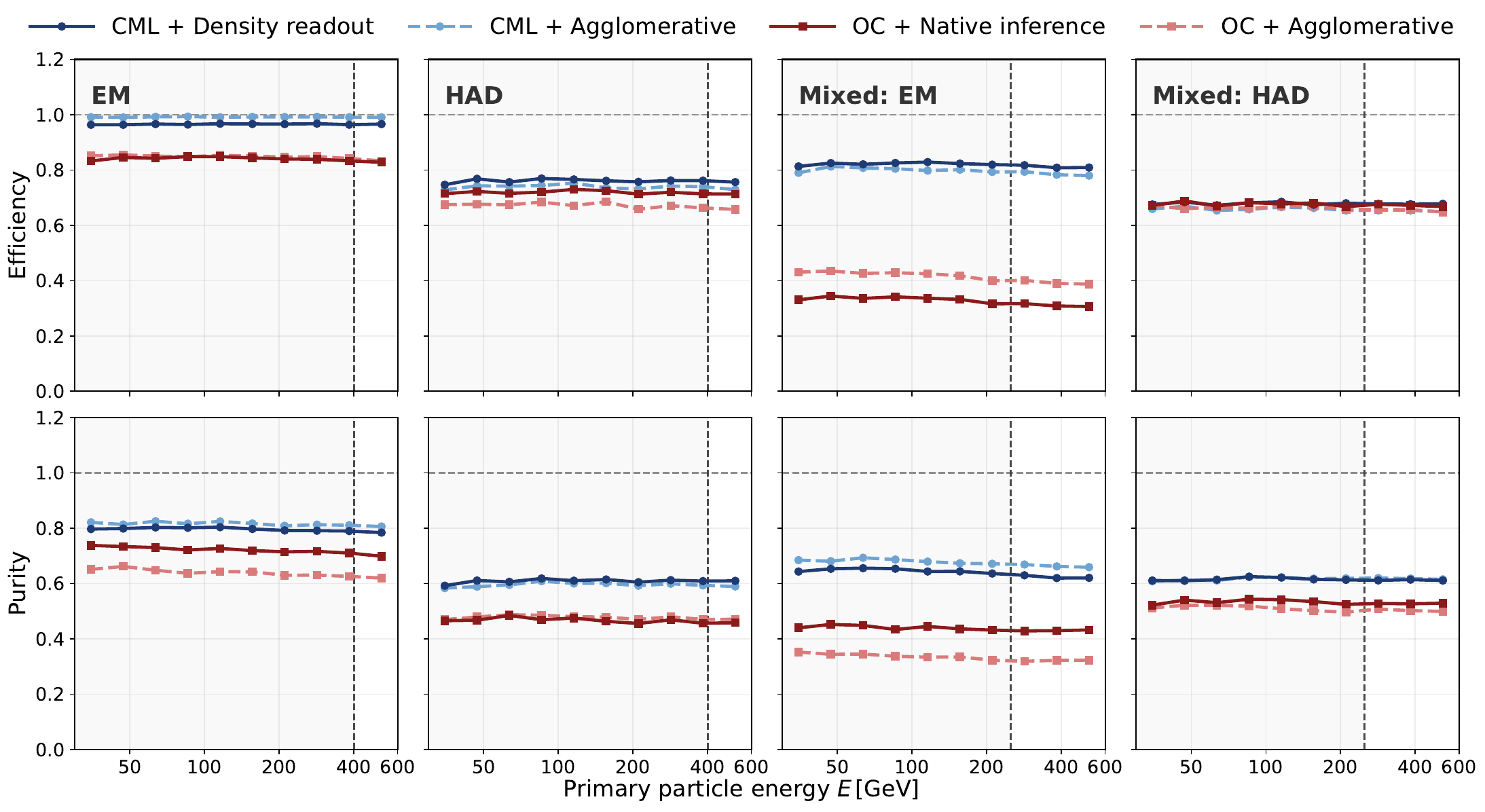}
    \caption{Reconstruction efficiency (top) and purity (bottom) as functions of primary particle energy for electromagnetic (EM) and hadronic (HAD) showers. Columns show models trained on EM, trained on HAD, and a mixed-trained model evaluated separately on EM and HAD showers. The vertical dashed line indicates the upper boundary of the training energy range. The dominant differences are largely independent of energy, indicating that performance is controlled primarily by the learned representation rather than by the absolute shower energy.}
    \label{fig:Metrics_Enegry}
\end{figure*}

\paragraph{Energy dependence.}

Figure~\ref{fig:Metrics_Enegry} shows reconstruction efficiency and purity as a function of the primary particle energy. For the dedicated EM model, CML achieves near-perfect efficiency (${\sim}99\%$) and substantially higher purity ($\sim$80--82\%) than OC, which remains below $85\%$ efficiency and $72\%$ purity. For HAD showers, efficiencies are more similar across methods ($\sim$70--76\%), but CML retains a clear purity advantage of $\sim$10--15\%.

The mixed-trained model reveals a stronger separation. For EM showers, OC degrades severely, with efficiencies dropping to $\sim$30--40\%, whereas CML retains efficiencies above $80\%$ and substantially higher purity. For HAD showers, efficiencies remain comparable, but CML again achieves higher purity.

Across all cases, performance remains stable beyond the training range, indicating that the dominant differences arise from the learned representation rather than from any strong dependence on primary energy. This is consistent with the embedding geometry, for which the separation margin is controlled mainly by local shower structure rather than absolute shower energy.

\begin{figure*}[t]
    \centering
    \includegraphics[width=1\linewidth]{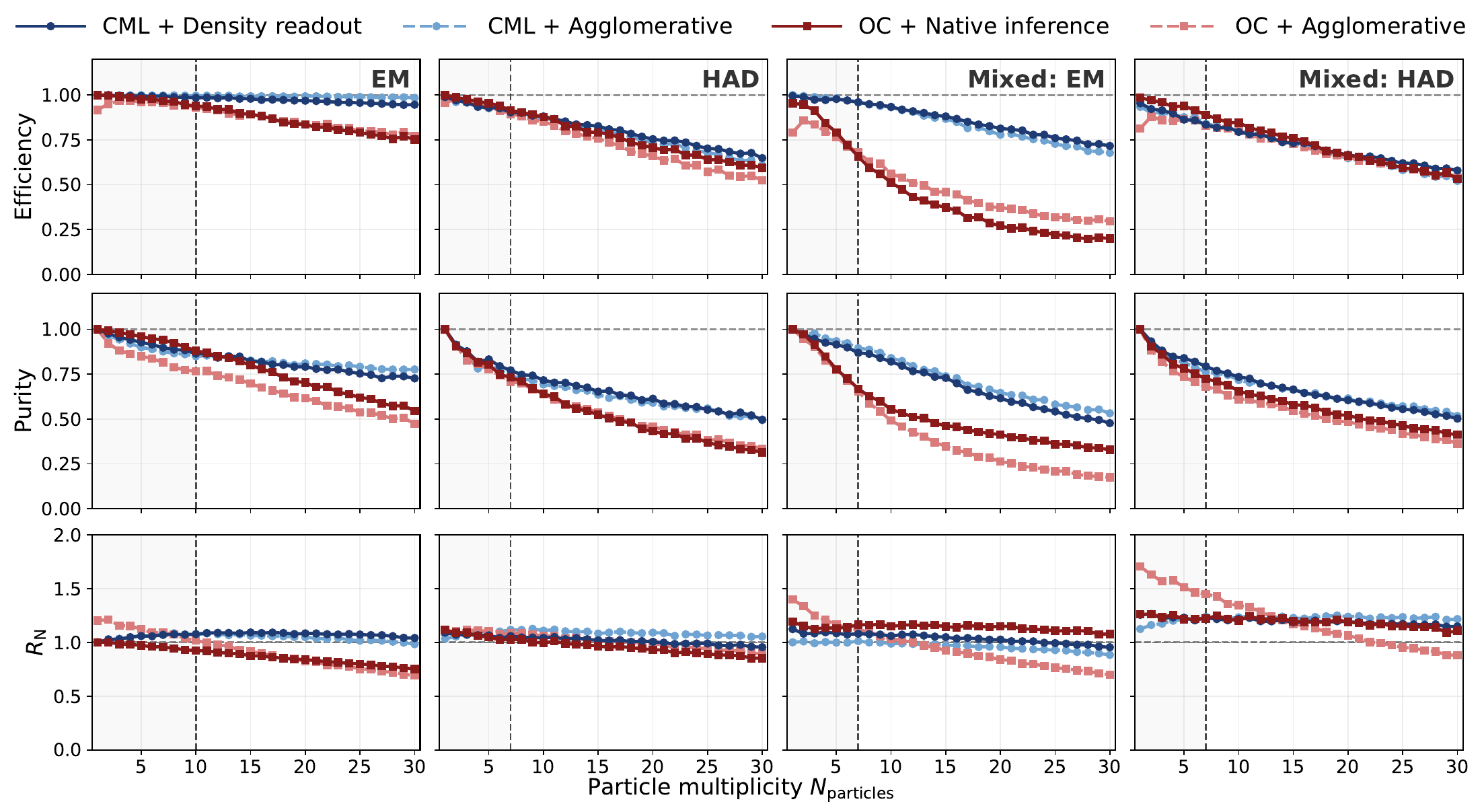}
    \caption{Reconstruction efficiency (top), purity (middle), and number ratio (bottom) as functions of particle multiplicity $N$ for electromagnetic (EM) and hadronic (HAD) showers. Columns show models trained on EM, trained on HAD, and a mixed-trained model evaluated separately on EM and HAD showers. The vertical dashed line indicates the upper boundary of the training multiplicity range. The performance gap between CML and OC increases strongly with multiplicity, showing that clustering stability in dense environments is determined by the underlying embedding geometry.}
    \label{fig:Metrics_Multiplicity}
\end{figure*}

\paragraph{Multiplicity dependence.}

Figure~\ref{fig:Metrics_Multiplicity} shows performance as a function of the particle multiplicity. This is the regime in which the separation between methods is most pronounced.

At low multiplicity, all methods perform similarly. As multiplicity increases, clear differences emerge. For EM showers at $N=30$, CML maintains high efficiency ($\sim$95--98\%) and purity ($\sim$73--78\%), while OC degrades to $\sim$75\% efficiency and $\sim$47--55\% purity. The number ratio remains close to unity for CML but drops significantly for OC, indicating substantial merging.

A similar trend is observed for HAD showers. At $N=30$, CML improves purity by nearly $20$ percentage points over OC while also providing a modest but systematic efficiency gain. Again, CML maintains a more stable $R_{N}$, indicating improved control of merging in dense environments.

The mixed-trained model shows the strongest contrast. For EM showers at $N=30$, CML remains functional with efficiencies of $\sim$70\%, whereas OC collapses to $\sim$20--30\%. For HAD showers, efficiencies remain comparable, but CML retains a clear purity advantage.

These trends can be understood directly from the embedding margins. For CML, the margin distributions remain narrow and concentrated near the separation boundary across multiplicities, so a single clustering scale continues to separate showers even as overlap increases. For OC, the behaviour depends strongly on shower type. In EM events, the margin becomes extremely broad, so no global threshold can separate showers consistently; this explains the large degradation in $R_{N}$ and purity, particularly for agglomerative clustering. In HAD events, the OC margin is less broad but systematically negative, so clustering is somewhat more stable but still intrinsically ambiguous, leading to persistent mis-clustering and reduced purity. The mixed setting makes this contrast most explicit: CML produces similar margin distributions for EM and HAD showers and therefore supports a common threshold, whereas OC does not, causing the EM performance of the mixed model to collapse while the HAD performance remains closer to that of the HAD-only model.

\begin{figure*}[t]
    \centering
    \includegraphics[width=1\linewidth]{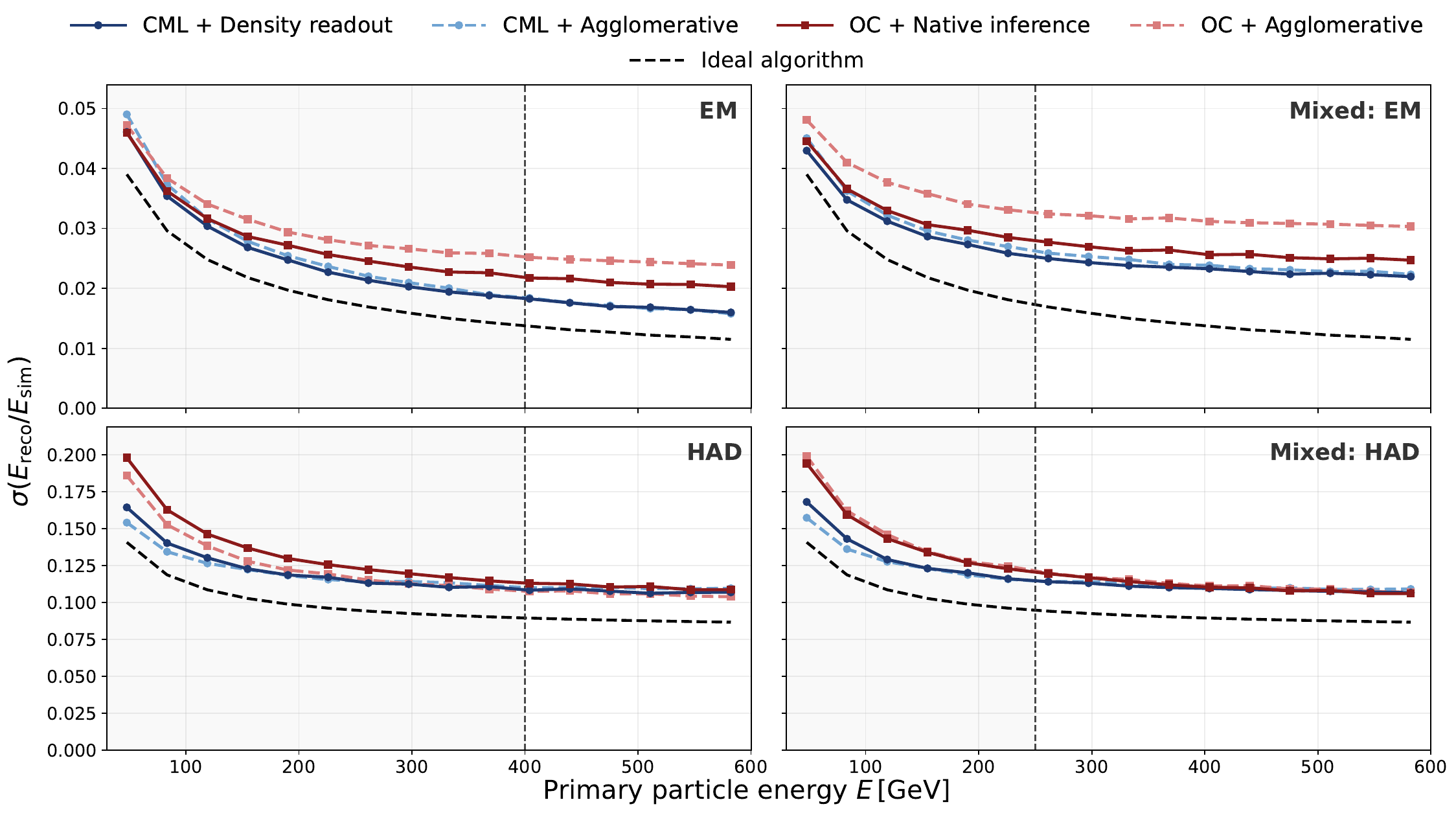}
    \caption{Energy resolution as a function of primary particle energy for electromagnetic (EM) and hadronic (HAD) showers. Panels show models trained on EM, a mixed-trained model evaluated on EM, trained on HAD, and a mixed-trained model evaluated on HAD. The vertical dashed line marks the upper boundary of the training energy range. The black dashed curve denotes the ideal pattern-recognition limit, defined as the resolution obtained by a perfect clustering algorithm with no merging or splitting, and represents a lower bound on the achievable resolution given the detector response. Improvements in CML resolution follow directly from its improved clustering purity and reduced merging. Results are averaged over the full test dataset spanning multiplicities from 1 to 30 particles per event.}
    \label{fig:Metrics_Resolutions}
\end{figure*}

\paragraph{Energy resolution.}

Figure~\ref{fig:Metrics_Resolutions} shows the corresponding energy resolution. For EM showers, CML consistently achieves the best resolution and remains closest to the ideal pattern-recognition limit. At high energy, $\mathcal{O}$(600\,GeV), CML reaches $\sim$1.6\%, compared with $\sim$2.0\% for OC with native inference and $\sim$2.4\% for OC with agglomerative clustering.

For HAD showers, the separation is smaller but remains systematic. The CML method provides the largest improvement at low energy, where the resolution is dominated by clustering quality rather than intrinsic detector response (e.g. $15.4\%$ versus $18$--$20\%$ at 50\,GeV), while at higher energies the curves approach a constant term and the gap narrows.

These results are the direct consequence of the clustering behaviour. The stable separation margins of CML reduce both merging and fragmentation, leading to more accurate assignment of energy deposits to reconstructed objects. In contrast, the broad and negative margins of OC lead to systematic mis-assignment of energy, particularly in dense environments, and therefore to degraded resolution.

\paragraph{Robustness to simulation variations.}

\begin{figure}[t]
    \centering
    \includegraphics[width=1.0\linewidth]{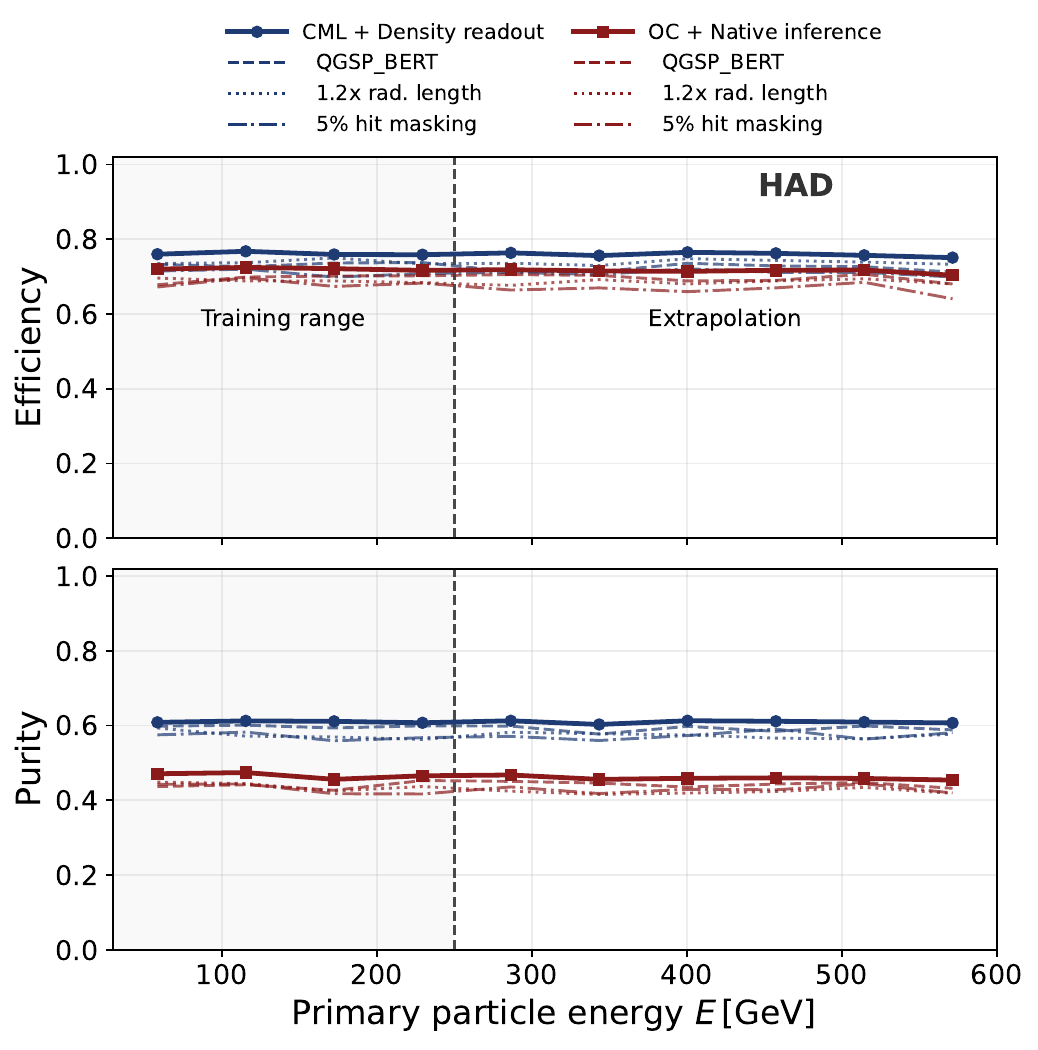}
    \caption{Reconstruction efficiency (top) and purity (bottom) for the HAD model as functions of primary particle energy under three systematic variations: $1.2\times$ hadronic interaction length scaling, QGSP\_BERT physics list, and $5\%$ random hit masking. The nominal performance is shown for reference. Neither method shows significant degradation, and the performance gap between CML and OC is preserved under all variations.}
    \label{fig:robustness}
\end{figure}

To assess the sensitivity of both methods to simulation mis-modelling, we evaluate the HAD model under three systematic variations: a $1.2\times$ scaling of the hadronic interaction length, a change of Geant4 physics list from FTFP\_BERT to QGSP\_BERT, and random masking of $5\%$ of hits. Figure~\ref{fig:robustness} shows reconstruction efficiency and purity for CML and OC under these variations. In all cases, both methods show only modest degradation relative to the nominal, and the performance gap between CML and OC is preserved across all tested variations. This indicates that the relative advantage of CML is stable under the types of simulation mis-modelling considered here.

\paragraph{Dependence on embedding dimensionality.}

Table~\ref{tab:mixed_results} shows the mean reconstruction performance of the mixed-trained model for 16- and 4-dimensional embeddings, evaluated on an independent mixed sample. Reducing the embedding dimension leads to a modest degradation for all methods, but the relative ordering remains unchanged: CML continues to outperform OC in both purity and efficiency across all configurations. This shows that the observed advantage is not specific to the 16-dimensional setting. The dimension of 16 was selected following a preliminary sweep over 4, 16, 128, and 512 dimensions; performance saturates at 16 for both methods, with degradation at higher dimensionalities consistent with distance concentration in high-dimensional spaces. The 4-dimensional result is included as a lower bound relevant to OC-based approaches, where lower-dimensional clustering coordinates are common.

\begin{table}[htbp]
\centering
\caption{Mean reconstruction purity and efficiency for the mixed-trained model evaluated on the combined EM and HAD test sample for 16- and 4-dimensional embeddings. The relative ordering between methods is unchanged when the embedding dimensionality is reduced, showing that the CML advantage is not driven by latent-space size alone.}
\label{tab:mixed_results}

\begin{tabular}{llccc}
\toprule
Model & Dimension & Clustering & Purity & Efficiency \\
\midrule

\multirow{4}{*}{CML}
    & 16 & Agglomerative & 0.722 & 0.932 \\
    & 16 & Density       & 0.710 & 0.945 \\
    & 4  & Agglomerative & 0.690 & 0.887 \\
    & 4  & Density       & 0.637 & 0.905 \\

\midrule

\multirow{4}{*}{OC}
    & 16 & Agglomerative & 0.469 & 0.849 \\
    & 16 & Density       & 0.520 & 0.838 \\
    & 4  & Agglomerative & 0.512 & 0.813 \\
    & 4  & Density       & 0.500 & 0.866 \\

\bottomrule
\end{tabular}
\end{table}

\section{Discussion and Conclusions}

We have presented a clustering framework for point-cloud segmentation in high-granularity calorimeters based on CML. Rather than learning object-centric clustering variables, the method learns a representation in which hits from the same shower are placed nearby and hits from different showers are separated, with clustering applied only as a readout of the learned geometry. This decoupling allows the representation to be optimized for pairwise compatibility while retaining flexibility in the choice of inference procedure.

The central result of this work is that the learned embedding geometry directly determines clustering performance. The CML approach produces narrow separation-margin distributions that remain positive for EM showers and only slightly negative for HAD showers, indicating a stable and well-defined clustering scale even in dense environments. 
In contrast, OC yields substantially broader and often negative margin distributions, particularly for EM showers, implying strong overlap between intra- and inter-shower distances and therefore intrinsically ambiguous clustering decisions. 
These geometric differences explain the observed reconstruction behaviour: CML consistently achieves higher purity, higher efficiency, more stable $R_{N}$, and improved energy resolution, with the largest gains appearing at high multiplicity where shower overlap is most severe.

The mixed-training results provide the strongest evidence for stability across heterogeneous particle types. The CML method maintains similar separation scales for EM and HAD showers, allowing a single clustering threshold to operate effectively across both particle types. 
By contrast, OC learns different geometric structure for EM and HAD showers. This leads to a pronounced degradation for EM showers in particular, indicating that the object-centric formulation is less able to accommodate heterogeneous shower topologies within a single model.

Taken together, these results show that, for highly granular calorimeter reconstruction, learning a stable similarity geometry is more effective than learning explicit object-centric clustering variables. More broadly, they suggest that contrastive metric learning provides a robust alternative for dense point-cloud segmentation problems in which object boundaries are ambiguous, overlap is common, and inference must remain stable under changing event complexity. We propose to test this strategy with ultra-realistic simulations of the HGCAL detector in high-pileup conditions, as implemented in the CMS software stack.

\section*{Acknowledgements}

We thank Sunanda Banerjee for the help in creating a realistic dataset corresponding to particle showers in a highly granular detector. The neural networks in this study have been trained on the Imperial College RCS HPC cluster and the Oscar cluster at Brown University. L. G. and L. N. are supported by the DOE, Office of Science, Office of High Energy Physics Early Career Research program under Award No. DE-SC0026288. B. M. acknowledges the support of Schmidt Sciences.

\FloatBarrier
\bibliography{ref}
\end{document}